\documentclass[12pt]{article}

\usepackage{amsmath,amssymb}
\usepackage{graphicx,a4,psfrag,here,epsfig,epsf}
\usepackage{pstricks}
\usepackage{pst-node}
\usepackage{citesort}
\allowdisplaybreaks[1]

\begin{document}
\newcommand{\bs}{\boldsymbol}
\newcommand{\sh}{\not\!}
\newcommand{\lag}{{\mathcal L}}
\newcommand{\ol}{\overline}
\newcommand{\co}{\; \; ,}
\newcommand{\per}{ \; .}
\newcommand{\nn}{\nonumber}

\newcommand{\lt}{\left}
\newcommand{\rt}{\right}
\renewcommand{\d}{\partial}
\newcommand{\fc}{\frac}

\newcommand{\rR}{R}
\newcommand{\rr}{{\cal R}}
\newcommand{\mc}{\multicolumn{2}{|c|}}
\newcommand{\mcc}{\multicolumn{3}{|c|}}
\newcommand{\nl}{\!\! & \!\!}
\newcommand{\av}[1]{\langle {#1}\rangle}

\def\beq{\begin{equation}}
\def\eeq{\end{equation}}
\def\bea{\begin{eqnarray}}
\def\eea{\end{eqnarray}}
\def\eq{\begin{eqnarray}}
\def\en{\end{eqnarray}}
\newcommand{\bes}{\begin{split}}
\newcommand{\ees}{\end{split}}
\newcommand{\bed}{\begin{displaymath}}
\newcommand{\eed}{\end{displaymath}}
\def\prop{{\mathcal D}}

\newcommand{\mpi}{M_\pi^2}
\newcommand{\mk}{M_K^2}
\newcommand{\me}{M_\eta^2}
\newcommand{\equ}{\,=\,}
\newcommand{\no}{\nonumber}
\newcommand{\noh}{\nonumber \hskip -10cm}
\newcommand{\Order}{\mathcal{O}}
\newcommand{\Lagr}{\mathcal{L}}
\newcommand{\M}{\mathcal{M}}
\numberwithin{equation}{section}
\newcommand{\ed}{\end{document}}
\newcommand{\scs}{\, , \,}
\newcommand{\sem}{\, ; \,}
\newcommand{\nnnl}{\nonumber\\}
\newcommand{\fs}{\, . \,}
\newcommand{\be}{\begin{eqnarray}}
\newcommand{\ee}{\end{eqnarray}}
\newcommand{\bra}{\right\rangle}
\newcommand{\bla}{\left\langle}
\newcommand{\word}[1]{{\mbox{{#1}\,}}}
\newcommand{\words}[1]{{\mbox{\small{#1}}}}
\newcommand{\wordt}[1]{{\mbox{\tiny{#1}}}}

\def\query#1{\marginpar{\begin{flushleft}\footnotesize#1\end{flushleft}}}%

\begin{titlepage}

\vspace{1cm}
\begin{center}{\Large\bf A framework for the calculation of the
\boldmath $\Delta N\gamma^*$ transition form factors on the lattice}

\vspace{0.5cm}
\today

\vspace{0.5cm}
Andria Agadjanov$^{a,b}$, V\'eronique Bernard$^c$, Ulf-G. Mei{\ss}ner$^{a,d}$
and Akaki Rusetsky$^a$

\vspace{2em}
\footnotesize{\begin{tabular}{c}
$^a\,$ Helmholtz-Institut f\"ur Strahlen- und Kernphysik (Theorie) and\\
Bethe Center for Theoretical Physics,\\
 $\hspace{2mm}$   Universit\"at Bonn, D-53115 Bonn, Germany\\[2mm]
$^b\,$ St. Andrew the First-Called Georgian University of the Patriarchate of Georgia,\\ Chavchavadze Ave. 53a, 0162, Tbilisi, Georgia\\[2mm]
$^c\,$Institut de Physique Nucl\'eaire, CNRS/Univ. Paris-Sud 11 (UMR 8608),\\
F-91406 Orsay Cedex, France\\[2mm]
$^d\,$ 
Institute for Advanced Simulation (IAS-4), Institut f\"ur Kernphysik 
(IKP-3) and\\ J\"ulich Center for Hadron Physics,
Forschungszentrum J\"ulich, D-52425 J\"ulich, Germany
\end{tabular}  }

\vspace{1cm}

\begin{abstract}

\noindent
Using the non-relativistic effective field theory framework in a finite volume,
we discuss the extraction of the $\Delta N\gamma^*$ transition form factors 
from lattice data. A counterpart of the L\"uscher approach for the {\em matrix
  elements} of unstable states is formulated. In particular, we thoroughly
discuss various kinematic settings, which are used in the calculation 
of the above matrix
element on the lattice. The emerging L\"uscher-Lellouch factor and the analytic
continuation of the matrix elements into the complex plane are also 
considered in detail. A full
group-theoretical analysis of the problem is made, including the 
partial-wave mixing and projecting out the invariant form factors from
data.
\end{abstract}

\vspace{1cm}
\footnotesize{\begin{tabular}{ll}
{\bf{Pacs:}}$\!\!\!\!$& 12.38.Gc, 13.40.Hq, 13.75.Gx
\\
{\bf{Keywords:}}$\!\!\!\!$& Lattice QCD, transition form factors,
non-relativistic EFT, L\"uscher equation
\\
\end{tabular}}
\end{center}
\end{titlepage}

\setcounter{page}{2}

\section{Introduction}
\label{sec:intro}

In recent years, the calculation of the $\Delta N\gamma^*$ transition 
form factors on the lattice has been carried out, see Refs.~\cite{Alexandrou:2010uk,Alexandrou:2011ga,Alexandrou:2007dt}. 
The electromagnetic, 
axial and pseudoscalar form factors of the $\Delta$-resonance have been also 
studied~\cite{Alexandrou:2008bn,Alexandrou:2013opa}. It should be noted,
however, that in these simulations the quark mass values are large enough
so that the $\Delta$ is a stable
particle and thus using the standard formalism for the analysis of the lattice
data on these form factors is justified. 
On the other hand, lattice simulations with physical
quark masses have already been performed. At such quark masses, the $\Delta$ is
not stable anymore and the data should be analyzed properly to extract the
parameters of the resonance (see, e.g.~\cite{Alexandrou:2013ata}).

It is well known that  resonances can not be identified with isolated energy levels
in lattice QCD simulations which are necessarily performed in a finite volume.
In order to determine the mass
and width from the measured spectrum, one first extracts the scattering
phase shift by using the L\"uscher equation~\cite{Luescher-torus}. At the next
step, using some parameterization for the $K$-matrix (e.g., the
effective-range expansion), a continuation 
into the complex energy plane is
performed. Resonances correspond to the poles of the scattering $T$-matrix
on the second Riemann sheet, and the real and imaginary parts of the pole
position define the mass and the width of a resonance. 
 This is a pretty standard
procedure that has been used in a number of recent
papers~\cite{Aoki-rho,Schierholz-rho,Mohler-resonances}. 
A generalization of the approach to moving frames has been first proposed in
Ref.~\cite{Rummukainen}, and a full
group-theoretical analysis of the L\"uscher equation in moving frames, 
including the issues related to the scattering
of the spin-non-zero particles, has been carried out, e.g., in
Refs.~\cite{Fu,Lage-distributions,Luu,Leskovec,Schierholz-group,LiLiu,Briceno}.
An alternative, albeit
a closely related procedure consists in fitting the data to the energy
spectrum by using unitarized ChPT in a finite
volume~\cite{oset,Doring-scalar,oset-in-a-finite-volume}. The above approach qualitatively
amounts to a parameterization of the $K$-matrix through the solution of the
equations of the unitarized ChPT (in the infinite volume) that may {\em a priori}
have a larger range of applicability than the effective-range expansion. Note
also that the approach has been generalized to the multichannel scattering
case~\cite{Lage-KN,Lage-scalar,He,Sharpe,Briceno-multi,Liu,PengGuo-multi}. An analysis of the
two-channel case has been carried already out for the toy model in 1+1
dimensions~\cite{PengGuo-ising} and should now be applied in physically
interesting cases. Further, using twisted boundary
conditions~\cite{Bedaque,Sachrajda,rest-twisted} to 
facilitate the accurate extraction of the resonance parameters has been
advocated, e.g., in Refs.~\cite{Lage-scalar,oset,Doring-scalar}, and the possibility of
the partial twisting has been investigated in
Refs.~\cite{Sachrajda,Chen,Agadjanov-twisted}. Last but not least, recently an
extension of the L\"uscher approach to the 3-particle case has been proposed
by several groups~\cite{Polejaeva,PengGuo-3,Briceno-3,Sharpe-3}, albeit there
is still much more work required in this direction.

As one sees from the above discussion, up to date
the framework for the extraction of the
resonance parameters (the mass and the width) 
from lattice data is well established (at
least, for the resonances that do not decay into three or more particles). For
the calculation of the more complicated quantities, e.g., the resonance
form factors, the old approach that treats the resonance as a stable state, is
still widely used, albeit it is clear that the same problems arise also here.
What one needs is a {\em generalization of the L\"uscher 
finite-volume approach
to the form factors.} 
In our recent papers~\cite{Hoja1,Hoja2}, we have formulated such a
generalization, considering the 
form factor of a spinless resonance. 
The procedure of extracting the resonance form factor from the {\em form factors
  of the eigenstates of the Hamiltonian}, which are actually measured on the
lattice, closely resembles the procedure of extracting the resonance pole
position and also implies the analytic continuation into the complex energy
plane by using, e.g., the effective range expansion. There are, however, 
differences as well. Most notably, as was shown in Ref.~\cite{Hoja2}, in 3+1
dimensions, due to the presence of the so-called finite fixed points, the procedure of
the analytic continuation and taking the infinite-volume limit is no more
straightforward and measuring the form factor for at least two different energy
levels is required, in order to achieve an unambiguous extraction of the
resonance form factor. 

The present paper is a continuation and the generalization of
Refs.~\cite{Hoja1,Hoja2} in two aspects:
\begin{itemize}
\item[i)]
In Refs.~\cite{Hoja1,Hoja2}, an elastic resonance form factor (an example:
the electromagnetic form factor $\Delta\Delta\gamma^*$) has been considered. In
this paper, we address transition form factors, in particular,
$\Delta N\gamma^*$ that is more interesting from the phenomenological point
of view. It turns out that the presence of one stable particle in the 
out-state (here, the nucleon) leads to crucial simplifications. As we shall
demonstrate, the finite fixed points do not exist in this case, so the measurement of
a single energy level (albeit at several volumes) 
will suffice. We shall discuss in detail
various lattice settings, which provide an access to the measurement of the
form factor.
\item[ii)]
All particles and currents, considered in Refs.~\cite{Hoja1,Hoja2}, were
scalar. On the other hand, the particles, whose form factors we want to
calculate, have spin. In this paper, we consider the inclusion of spin into
the formalism and carry out a full group-theoretical analysis of the obtained
equations along the lines described in Ref.~\cite{Schierholz-group}.
\end{itemize}

The paper is organized as follows. In section~\ref{sec:infinite}, 
we start from defining the resonance
form factor in the infinite volume and discuss the analytic continuation into
the complex plane. The projection of various scalar form factors will be
considered. In section~\ref{sec:setting} we consider the kinematics,
which should be used
for measuring the matrix elements of a current between the eigenstates of the
Hamiltonian. This issue is very important for performing the analytic
continuation of the above matrix element, {\em keeping the relative three-momentum of the
  photon and nucleon fixed}. Further, in section~\ref{sec:diagrams}, we calculate these
matrix elements within the non-relativistic effective field theory (EFT) and
demonstrate that the finite fixed points are absent, when one of the external
particles is stable. The initial-state interactions, which manifest themselves
in the L\"uscher-Lellouch factor, should be properly included in order to take
into account the difference in normalization of the matrix elements in the
infinite and in a finite volume. In section~\ref{sec:prescription} we
collect all bits and pieces and formulate 
a prescription for the extraction of the
resonance transition form factor from data. The section~\ref{sec:concl}
contains our conclusions. Finally, in the Appendix, the formulae for
the partial-wave expansion of the photoproduction amplitudes are displayed.

\section{Resonance form factor in the infinite volume}
\label{sec:infinite}     

To the best of our knowledge, the procedure for calculating the matrix 
elements of operators between the bound state vectors in field theory
has been first addressed
in Ref.~\cite{Mandelstam} (a very detailed and transparent discussion of the
problem can be found in Ref.~\cite{Huang-Weldon}). In short, the procedure
boils down to the following. For simplicity, consider the scalar case first.
Let $|P\rangle$ be a stable bound state moving 
with a four-momentum $P_\mu$ with $P_\mu P^\mu=M_B^2$. The fact that this is a
bound state and not an elementary state is equivalent to the statement that
$\langle 0|\phi(x)|P\rangle=0$, where $\phi(x)$ stands for any field which
is present in the Lagrangian. Consider now an operator $O(X)$ built from the
elementary fields $\phi$. The operator $O(X)$ can be either local or non-local. 
In the latter case, $X$ denotes the center-of-mass coordinate of the fields 
entering in $O(X)$. The only requirement on this operator is that 
$\langle 0|O(X)|P\rangle\neq 0$. The Fourier-transform of the two-point
function of the  operators $O$ has a pole at $P^2=M_B^2$:
\eq
i\int d^4Xe^{iPX}\langle 0|TO(X) \bar O(0)|0\rangle=\frac{Z_B}{M_B^2-P^2}+
\mbox{regular terms at $P^2\to M_B^2$,}
\en
where $Z_B$ is the wave function renormalization constant of the bound state
(for the scalar operators, considered here, the conjugated operator $\bar O=O^\dagger$).

Let us now consider the three-point function with any local operator $J$ 
(the ``current''). This function has a double pole
\eq
F(P,Q)&=&i^2\int d^4Xd^4Ye^{iPX-iQY}\langle 0|TO(X)J(0)\bar O(Y)|0\rangle
\nonumber\\[2mm]
&=&\frac{Z_B^{1/2}}{M_B^2-P^2}\,\langle P|J(0)|Q\rangle\,
\frac{Z_B^{1/2}}{M_B^2-Q^2}+\cdots\, ,
\en
where the ellipses stand for the less singular terms. From this equation 
one immediately sees that the matrix element of a current $J$ between the 
bound state vectors is defined through
\eq\label{eq:matrixelement-bound}
\langle P|J(0)|Q\rangle =\lim_{P^2,Q^2\to M_B^2}Z_B^{-1}(M_B^2-P^2)(M_B^2-Q^2)F(P,Q)\, .
\en
Due to the Lorenz-invariance, the matrix element in the l.h.s. of 
this equation is a function of a single scalar variable $t=(P-Q)^2$.

Note also that 
the expression given in Eq.~(\ref{eq:matrixelement-bound}) defines the matrix
element between bound states. In a completely similar manner, it is possible
to define the matrix elements between a bound state and an elementary state,
or between two different bound states -- all differences boil down to the
proper choice of the operators $O$.

In case of a {\em resonance} rather than a bound system, no corresponding
single-particle state exists in the Fock space. In the literature, one 
encounters two different approaches to the problem. One approach, which 
implies the definition of the form factor from the amplitudes measured 
at {\em real energies,} invokes the Breit-Wigner parameterization of the 
resonant amplitude and extracts the resonance form factors at the energy where
the scattering phase shift passes through $90^{\sf o}$. Within the second approach,
the resonance form factor is defined through the continuation to the resonance
pole position in the complex plane, as described below.
Let $O(X)$ be the operator with the quantum numbers of a resonance. 
The two-point function of the operators $O$ develops a pole on the unphysical 
Riemann sheet in the complex plane
\eq
i\int d^4Xe^{iPX}\langle 0|TO(X) \bar O(0)|0\rangle=\frac{Z_R}{s_R-P^2}+
\mbox{regular terms at $P^2\to s_R$,}
\en
where the quantities $s_R,Z_R$ are now complex. The real and imaginary parts of
$E_R=\sqrt{s_R}$ give the mass and the half-width of the resonance, respectively.

Further, the three-point function develops a double 
pole in the complex plane, and
the {\em resonance matrix element} of any current $J$ is still defined
by a formula similar to the Eq.~(\ref{eq:matrixelement-bound}):
\eq\label{eq:matrixelement-resonance}
\langle P|J(0)|Q\rangle =\lim_{P^2,Q^2\to s_R}Z_R^{-1}(s_R-P^2)(s_R-Q^2)F(P,Q)\, .
\en
We would like to stress that the quantity on the l.h.s. of 
Eq.~(\ref{eq:matrixelement-resonance}) is a mere notation for the matrix 
element: there exists no isolated resonance state $|P\rangle$ in the spectrum.
Again, due to the Lorentz-invariance, this quantity is a function of a 
single variable $t=(P-Q)^2$.

The following questions arise naturally in connection to the procedure
described above:
\begin{itemize}
\item[i)]
Is it not possible to avoid the analytic continuation into the complex energy
plane?

\item[ii)]
Experiments can only be performed for real energies. How does one 
perform the analytic continuation of the experimental data?

\end{itemize} 

In brief, answers to these question are:

\begin{itemize}

\item[i)]
Relating the form factor to the measured scattering amplitudes by using, e.g.,
the Breit-Wigner parameterization, yields a model-dependent result, since
the background is not known. Consequently, the form factor, extracted at
the real energies, will be process-dependent. This problem does not arise,
when an analytic continuation to the resonance pole is performed.
The resonance matrix elements extracted through the analytic continuation,
are the quantities that characterize the resonance
itself and not the process where they were determined.

\item[ii)]
The analytic continuation of the experimental data (e.g., in order to extract
the magnetic moment of a $\Delta$-resonance) is, in general, a very difficult
procedure and is severely limited by the experimental uncertainties. However,
 the goal is still worth trying, see the arguments above.

\end{itemize}

It should be mentioned that both
definitions of the form factor: 
on the real axis (see, e.g.,~\cite{Aznauryan,Drechsel})
as well as at the resonance pole~\cite{Sarantsev}, have been already used for the analysis 
of the experimental data
(the latter work contains also the comparison of the resonance parameters,
 extracted by using different methods). In order to make it possible to
compare lattice calculations with all existing experimental results, 
in this paper we provide the formulae which should be used on the real axis,
as well as in the complex energy plane. Here we stress once more  that 
only the definition, based on the analytic continuation, yields a 
resonance form factor that is devoid of any process-dependent ambiguities. 
Further, it will be explicitly demonstrated
that both methods yield the same result in the limit of the infinitely 
small width.

Up to now, all particles and operators considered were scalars. In order
to include the resonances of a generic spin, we follow closely the procedure
of Refs.~\cite{Gegelia}. Let $O_\alpha(X)$ be the interpolating field for a 
resonance. Here, $\alpha$ denotes the collection of indices characterizing a
resonance with spin (Dirac indices, vector indices). The two-point function
in the vicinity of the resonance pole has the following behavior:
\eq
i\int d^4Xe^{iPX}\langle 0|TO_\alpha(X) \bar O_\beta(0)|0\rangle
=\frac{Z_RP_{\alpha\beta}(P,s_R)}{s_R-P^2}+
\mbox{regular terms at $P^2\to s_R$,}
\en
where $P_{\alpha\beta}$ denotes the projector on the positive energy ``states''
\eq
P_{\alpha\beta}=\sum_\varepsilon u_\alpha(P,\varepsilon)\bar 
u_\beta(P,\varepsilon)\, ,\quad
\en
where the sum runs over spin projections on the third axis ($\varepsilon$) 
and  $u_\alpha(P,\varepsilon)$ denotes the solution of the free wave 
equation for a particle with a given spin.

Below, we shall give a construction of $u_\alpha(P,\varepsilon)$  
in case of the spin-1/2 and spin-3/2 particles. In case of the spin-1/2
particle, this quantity is given by~\cite{Gegelia}: 
\eq
u(P,1/2)&=&\sqrt{P^0+E_R}
\begin{pmatrix}
1\\[2mm] 0\\[2mm]\dfrac{P^3}{P^0+E_R}\\[4mm] \dfrac{P^1+iP^2}{P^0+E_R}
\end{pmatrix}\, ,
\nonumber\\[2mm]
u(P,-1/2)&=&\sqrt{P^0+E_R}
\begin{pmatrix}
0\\[2mm] 1\\[2mm]\dfrac{P^1-iP^2}{P^0+E_R}\\[4mm] \dfrac{-P^3}{P^0+E_R}
\end{pmatrix}\, ,
\nonumber\\[2mm]
\bar u(P,1/2)&=&\sqrt{P^0+E_R}
\begin{pmatrix}
1,& 0,&\dfrac{-P^3}{P^0+E_R},& \dfrac{-(P^1-iP^2)}{P^0+E_R}
\end{pmatrix}\, ,
\nonumber\\[2mm]
\bar u(P,-1/2)&=&\sqrt{P^0+E_R}
\begin{pmatrix}
0,& 1,&\dfrac{-(P^1+iP^2)}{P^0+E_R},& \dfrac{P^3}{P^0+E_R}
\end{pmatrix}\, .
\en
These spinors obey the Dirac equations with the complex ``mass'' $P^2=s_R$
\eq
(\not\!\! P-E_R) u(P,\varepsilon)=0\, ,\quad\quad
\bar u(P,\varepsilon)(\not\!\! P-E_R)=0
\en
as well as the identities
\eq
\sum_\varepsilon u(P,\varepsilon)\bar u(P,\varepsilon)
=(\not\!\! P+E_R)\, ,\quad\quad
 \bar u(P,\varepsilon) u(P,\varepsilon')=2E_R\delta_{\varepsilon\varepsilon'}\, .
\en
Note, however, that, if $E_R$ and $P_\mu$ are complex quantities, then, in general,
\eq
\bar u(P,\varepsilon)\neq u(P,\varepsilon)^\dagger\gamma_0\, .
\en
In case of a particle with a spin-3/2, one has to construct the solutions of the Rarita-Schwinger equation with a complex ``mass.'' To this end, we define three vectors ${\bf e}_\omega$ with $\omega=\pm 1,0$:
\eq
&&{\bf e}_{+1}=-\frac{1}{\sqrt{2}}\,
\begin{pmatrix}
1\\i\\0
\end{pmatrix}\, ,\quad\quad
{\bf e}_0=
\begin{pmatrix}
0\\0\\1
\end{pmatrix}\, ,\quad\quad
{\bf e}_{-1}=\frac{1}{\sqrt{2}}\,
\begin{pmatrix}
1\\-i\\0
\end{pmatrix}\, ,
\nonumber\\[2mm]
&&\bar{\bf e}_{+1}=-\frac{1}{\sqrt{2}}\,(1,-i,0)\, ,\quad\quad
\bar{\bf e}_0=(0,0,1)\, ,\quad\quad
\bar{\bf e}_{-1}=\frac{1}{\sqrt{2}}\,(1,i,0)\, .
\en
Further, define
\eq
f^\mu(P,\omega)&=&\biggl(\frac{{\bf e}_\omega\cdot {\bf P}}{E_R}\,,~
{\bf e}_\omega+\frac{{\bf P}({\bf e}_\omega\cdot{\bf P})}{E_R(P^0+E_R)}\biggr)\, ,
\nonumber\\[2mm]
\bar f^\mu(P,\omega)&=&\biggl(\frac{\bar {\bf e}_\omega\cdot {\bf P}}{E_R}\,,~
\bar{\bf e}_\omega+\frac{{\bf P}(\bar{\bf e}_\omega\cdot{\bf P})}{E_R(P^0+E_R)}\biggr)\, .
\en
The Rarita-Schwinger wave functions are given by group-theoretical expressions corresponding to the addition of spins 1 and 1/2:
\eq\label{eq:Rarita}
u^\mu(P,\lambda)&=&\sum_{\omega,\varepsilon}
\langle 1\,\omega\, 1/2\,\varepsilon|3/2\,\lambda\rangle f^\mu(P,\omega)u(P,\varepsilon)\, ,
\nonumber\\[2mm]
\bar u^\mu(P,\lambda)&=&\sum_{\omega,\varepsilon}
\langle 1\,\omega\,1/2\,\varepsilon|3/2\,\lambda\rangle 
\bar u(P,\varepsilon)\bar f^\mu(P,\omega)\, ,
\en
with $\langle \ldots\rangle$ the appropriate Clebsch-Gordan coefficients.
These wave functions obey the equations
\eq
&&(\not\!\!P-E_R)u^\mu(P,\lambda)=0\, ,\quad\quad 
P_\mu u^\mu(P,\lambda)=\gamma_\mu u^\mu(P,\lambda)=0\, ,
\nonumber\\[2mm]
&&\bar u^\mu(P,\lambda)(\not\!\!P-E_R)=0\, ,\quad\quad
\bar u^\mu(P,\lambda)P_\mu=\bar u^\mu(P,\lambda)\gamma_\mu=0\, ,
\en
and the identities
\eq
\sum_\lambda u^\mu(P,\lambda)\bar u^\nu(P,\lambda)
&=&-\frac{\not\!\!P+E_R}{2E_R}\biggl(g^{\mu\nu}-\frac{1}{3}\,\gamma^\mu\gamma^\nu
-\frac{2P^\mu P^\nu}{3E_R^2}+\frac{P^\mu\gamma^\nu-P^\nu\gamma^\mu}{3E_R}\biggr)\, ,
\nonumber\\[2mm]
\sum_\mu\bar u^\mu(P,\lambda) u_\mu(P,\lambda')&=&-2E_R\delta_{\lambda\lambda'}\, .
\en
Below, we shall restrict ourselves
to the case of the $\Delta N\gamma^*$ transition. However, the formalism can be directly generalized to particles with any spin. The three-point function
for this transition takes the form
\eq\label{eq:doublepole_NDg}
F^{\mu\rho}(P,Q)&=&i^2\int d^4Xd^4Ye^{iPX-iQY}\langle 0|TO^\mu(X)J^\rho(0)\bar\psi(Y)|0\rangle
\nonumber\\[2mm]
&=&\frac{Z_R^{1/2}}{s_R-P^2}\,\frac{Z_N^{1/2}}{m_N^2-Q^2}\,
\sum_{\lambda,\varepsilon}
u^\mu(P,\lambda)
\langle P,\lambda|J^\rho(0)|Q,\varepsilon\rangle\,
\bar u(Q,\varepsilon)
+\cdots\, .
\en
Here, $O^\mu(X)$ and $\psi(Y)$ denote the $\Delta$ and nucleon interpolating 
field operators, respectively, $J^\rho$ is the electromagnetic current,
$s_R$ is the $\Delta$-resonance pole position 
in the complex plane, and $m_N$ is the nucleon mass. 
The sum runs over the $\Delta$ and nucleon spin projections:
$\lambda=-3/2,-1/2,1/2,3/2$ and
$\varepsilon=-1/2,1/2$.
As already stated after Eq.~(\ref{eq:matrixelement-resonance}),
the matrix element that appears on the r.h.s. of the above equation is a mere
notation: there is no stable $\Delta$-state in the Fock space of the theory.
We shall use this notation throughout the paper.

Projecting out the matrix element from Eq.~(\ref{eq:doublepole_NDg}), we
get
\eq
\langle P,\lambda|J^\rho(0)|Q,\varepsilon\rangle&=&
\lim_{P^2\to s_R,~Q^2\to m_N^2}\frac{Z_R^{-1/2}}{2E_R}\,
\frac{Z_N^{-1/2}}{2m_N}\,
(s_R-P^2)(m_N^2-Q^2)
\nonumber\\[2mm]
&\times&\bar u_\mu(P,\lambda)F^{\mu\rho}(P,Q) u(Q,\varepsilon)\, .
\en
This matrix element can be expressed in terms of three scalar form factors
(see, e.g.~\cite{Scadron,Pascalutsa})
\eq\label{eq:scadron}
\langle P,\lambda|J^\rho(0)|Q,\varepsilon\rangle
=\biggl(\frac{2}{3}\biggr)^{1/2}\bar u_\mu(P,\lambda)\biggl\{
G_M(t){\cal K}_M^{\mu\rho}+G_E(t){\cal K}_E^{\mu\rho}
+G_C(t){\cal K}_C^{\mu\rho}\biggr\} u(Q,\varepsilon)\, ,
\en
where\footnote{For the Dirac matrices, we use the 
conventions of Ref.~\cite{IZ}.}
\eq\label{eq:scadron1}
{\cal K}_M^{\mu\rho}&=&
-\frac{3}{(E_R+m_N)^2-t}\,\frac{E_R+m_N}{2m_N}\,\epsilon^{\mu\rho\alpha\nu}
p_\alpha q_\nu\, ,
\nonumber\\[2mm]
{\cal K}_E^{\mu\rho}&=&
-{\cal K}_M^{\mu\rho}+6i\Delta^{-1}(t)\frac{E_R+m_N}{m_N}\,\gamma_5
\epsilon^{\mu\sigma\alpha\beta}p_\alpha q_\beta
\epsilon^{\rho\omega\nu\delta}g_{\sigma\omega}p_\nu q_\delta\, ,
\nonumber\\[2mm]
{\cal K}_C^{\mu\rho}&=&
3i\Delta^{-1}(t)\frac{E_R+m_N}{m_N}\,\gamma_5 q^\mu(q^2 p^\rho-(q\cdot p)q^\rho)\, ,
\en
and
\eq
p=\frac{1}{2}\,(P+Q)\, ,\quad
q=P-Q\, ,\quad
\Delta(t)=((E_R+m_N)^2-t)((E_R-m_N)^2-t)\, .
\en
$K_{M,E,C}$ are, respectively, the magnetic dipole, electric
quadrupole and Coulomb (longitudinal) quadrupole covariants.
In order to determine these three scalar form factors separately, it is
convenient to work in a special kinematics. We choose both 3-momenta ${\bf P}$
and ${\bf Q}$ along the third axis. Further, 
the formulae simplify considerably in 
the rest-frame of the resonance ${\bf P}=0$. Below, we shall adopt
this choice.

Using Eqs.~(\ref{eq:scadron}) and (\ref{eq:scadron1}), it is straightforward
to show that, in the rest-frame of the $\Delta$-resonance,
\eq\label{eq:proj}
\langle 1/2|J^3(0)|1/2\rangle
&=&i\,\frac{E_R-Q^0}{E_R}\,A\,G_C(t)\, ,
\nonumber\\[2mm]
\langle 1/2|J^+(0)|-1/2\rangle
&=&-i\sqrt{\frac{1}{2}}\,A
(G_M(t)-3G_E(t))\, ,
\nonumber\\[2mm]
\langle 3/2|J^+(0)|1/2\rangle
&=&-i\sqrt{\frac{3}{2}}A(G_M(t)+G_E(t))\, ,
\en
where
\eq\label{eq:A}
J^+(0)=\frac{J^1(0)+iJ^2(0)}{\sqrt{2}}\, , \quad\quad
A=\frac{E_R+m_N}{2m_N}\,\sqrt{2E_R(Q^0-m_N)}\, .
\en
Further, it can be shown that the following field operators
\eq\label{eq:O}
O_{3/2}(X)&=&\frac{1}{2}\,(1+\Sigma_3)\frac{1}{2}\,(1+\gamma_0)\frac{1}{\sqrt{2}}\,(O^1(X)-i\Sigma_3 O^2(X))\, ,
\nonumber\\[2mm]
O_{1/2}(X)&=&\frac{1}{2}\,(1-\Sigma_3)\frac{1}{2}\,(1+\gamma_0)\frac{1}{\sqrt{2}}\,(O^1(X)+i\Sigma_3 O^2(X))\, ,
\nonumber\\[2mm]
\tilde O_{1/2}(X)&=&\frac{1}{2}\,(1+\Sigma_3)\frac{1}{2}\,(1+\gamma_0)O^3(X)
\en
produce the $\Delta$-particles with spin projection $\lambda=3/2$
and $\lambda=1/2$, respectively. Here, $\Sigma_3$ denotes the 
$4\times 4$ matrix, describing the spin projection on the third axis.
In terms of the Pauli matrices, it is given by $\Sigma_3=\mbox{diag}(\sigma_3,\sigma_3)$. Note also that the operators
$O_{3/2}$ and $O_{1/2},\tilde O_{1/2}$ belong to the different irreducible
representations (irreps), $G_2$ and $G_1$, respectively,
 of the little group of the double cover of the
cubic group, corresponding to the boost momentum along the third 
axis~\cite{Schierholz-group}. Moreover, there are two different operators
that correspond to the spin projection $\lambda=1/2$, whereas there exists only
one operator for the projection $\lambda=3/2$, see Eqs. (113) and (114) of
Ref.~\cite{Schierholz-group}.

The field operators, projecting onto the states with a given 
third component of the nucleon, are constructed trivially:
\eq
\bar\psi_{\pm 1/2}(Y)=\bar\psi(Y)\frac{1}{2}\,(1\pm\Sigma_3)\frac{1}{2}\,(1+\gamma_0)\, .
\en
Using the above operators, we may construct the following three-point functions:
\eq
\tilde F_{1/2}(P,Q)
&=&i^2\int d^4Xd^4Ye^{iPX-iQY}\langle 0|T\tilde O_{1/2}(X)J^3(0)\bar\psi_{1/2}(Y)|0\rangle\, ,
\nonumber\\[2mm]
F_{1/2}(P,Q)
&=&i^2\int d^4Xd^4Ye^{iPX-iQY}\langle 0|TO_{1/2}(X)J^+(0)\bar\psi_{-1/2}(Y)|0\rangle\, ,
\nonumber\\[2mm]
F_{3/2}(P,Q)
&=&i^2\int d^4Xd^4Ye^{iPX-iQY}\langle 0|TO_{3/2}(X)J^+(0)\bar\psi_{1/2}(Y)|0\rangle\, .
\en
In the vicinity of the double pole, these functions behave as
\eq
\mbox{Tr}\,(\tilde F_{1/2}(P,Q))&=&i\biggl(\frac{2}{3}\biggr)^{1/2}
\frac{Z_R^{1/2}}{s_R-P^2}\,\frac{Z_N^{1/2}}{m_N^2-Q^2}\,
\frac{E_R-Q^0}{E_R}\,BG_C(t)+\cdots\, ,
\nonumber\\[2mm]
\mbox{Tr}\,(F_{1/2}(P,Q))&=&i\biggl(\frac{2}{3}\biggr)^{1/2}
\frac{Z_R^{1/2}}{s_R-P^2}\,\frac{Z_N^{1/2}}{m_N^2-Q^2}\,
\frac{1}{2}\,B(G_M(t)-3G_E(t))+\cdots\, ,
\nonumber\\[2mm]
\mbox{Tr}\,(F_{3/2}(P,Q))&=&i\biggl(\frac{2}{3}\biggr)^{1/2}
\frac{Z_R^{1/2}}{s_R-P^2}\,\frac{Z_N^{1/2}}{m_N^2-Q^2}\,
\frac{3}{2}\,B(G_M(t)+G_E(t))+\cdots\, ,
\en
where the trace is performed over the Dirac indices, and
\eq
B=\frac{E_R(E_R+m_N)}{m_N}\,|{\bf Q}|\, .
\en
So, with a special choice of the interpolating operators, the problem of a particle with spin boils down to the spinless case, considered in the beginning of this
section. The three form factors $G_C,G_M,G_E$ can be projected out individually.

\section{Extracting the form factors on the lattice}
\label{sec:setting}

Below, we adapt the formulae of the previous section and formulate the rules
for projecting out the form factors $G_C,G_M,G_E$ from the Euclidean Green 
functions on the lattice. Let us first restrict ourselves to the case when the 
$\Delta$ is stable and consider the following three-point functions at $t'>0,t<0$:  
\eq
\tilde R_{1/2}(t',t)
&=&\langle 0|\tilde {\cal O}_{1/2}(t')J^3(0)\bar{\cal \psi}^{\,\bf Q}_{1/2}(t)|0\rangle\, ,
\nonumber\\[2mm]
R_{1/2}(t',t)
&=&\langle 0|{\cal O}_{1/2}(t')J^+(0)\bar\psi^{\,\bf Q}_{-1/2}(t)|0\rangle\, ,
\nonumber\\[2mm]
R_{3/2}(t',t)
&=&\langle 0|{\cal O}_{3/2}(t')J^+(0)\bar\psi^{\,\bf Q}_{1/2}(t)|0\rangle\, ,
\en
where
\eq\label{eq:Xt}
\tilde {\cal O}_{1/2}(t')&=&\sum_{\bf X}\tilde O_{1/2}({\bf X},t')\, ,
\nonumber\\[2mm]
{\cal O}_{1/2}(t')&=&\sum_{\bf X} O_{1/2}({\bf X},t')\, ,
\nonumber\\[2mm]
{\cal O}_{3/2}(t')&=&\sum_{\bf X} O_{3/2}({\bf X},t')\, ,
\nonumber\\[2mm]
\bar\psi^{\,\bf Q}_{\pm 1/2}(t)&=&\sum_{\bf X}e^{i{\bf Q}{\bf X}}\bar\psi_{\pm 1/2}({\bf X},t)\, .
\en
The operators $\tilde {\cal O}_{1/2}(t'),{\cal O}_{1/2}(t'),{\cal O}_{3/2}(t')$
describe the $\Delta$ at rest, whereas $\bar\psi^{\,\bf Q}_{\pm 1/2}(t)$ 
corresponds to the nucleon moving with the 3-momentum ${\bf Q}$.
The operators on the r.h.s. of Eq.~(\ref{eq:Xt}) are given in Eq.~(\ref{eq:O})
with the substitution $\gamma_0\to\gamma_4$.

In the limit $t'\to+\infty$, $t\to-\infty$ only the one-particle $\Delta$ and 
nucleon states contribute:
\eq
\tilde R_{1/2}(t',t)
&\!\!\to\!\!&\frac{e^{-E_\Delta t'+E_Nt}}{4E_\Delta E_N}\,
\langle 0|\tilde {\cal O}_{1/2}(0)|1/2\rangle
\langle 1/2| J^3(0)|1/2\rangle
\langle 1/2|\bar{\cal \psi}^{\,\bf Q}_{1/2}(0)|0\rangle\, ,
\nonumber\\[2mm]
R_{1/2}(t',t)
&\!\!\to\!\!&\frac{e^{-E_\Delta t'+E_Nt}}{4E_\Delta E_N}\,
\langle 0|{\cal O}_{1/2}(0)|1/2\rangle
\langle 1/2|J^+(0)|-1/2\rangle
\langle -1/2|\bar\psi^{\,\bf Q}_{-1/2}(0)|0\rangle ,
\nonumber\\[2mm]
R_{3/2}(t',t)
&\!\!\to\!\!&\frac{e^{-E_\Delta t'+E_Nt}}{4E_\Delta E_N}\,
\langle 0|{\cal O}_{3/2}(0)|3/2\rangle
\langle 3/2| J^+(0)|1/2\rangle
\langle 1/2|\bar{\cal \psi}^{\,\bf Q}_{1/2}(0)|0\rangle\, ,
\en
where $E_\Delta=m_\Delta$ in the rest-frame of the $\Delta$ and $E_N=\sqrt{m_N^2+{\bf Q}^2}$ (here, $m_\Delta$ denotes the mass of a stable $\Delta$).
Further, we define the following 2-point functions
\eq
\tilde D_{1/2}(t)&=&\mbox{Tr}\,\langle 0|\tilde {\cal O}_{1/2}(t){\bar{\tilde {\cal O}}}_{1/2}(0)|0\rangle\, ,
\nonumber\\[2mm]
D_{1/2}(t)&=&\mbox{Tr}\,\langle 0|{\cal O}_{1/2}(t)\bar{\cal O}_{1/2}(0)|0\rangle\, ,
\nonumber\\[2mm]
D_{3/2}(t)&=&\mbox{Tr}\,\langle 0|{\cal O}_{3/2}(t)\bar{\cal O}_{3/2}(0)|0\rangle\, ,
\nonumber\\[2mm]
D^\pm_{\bf Q}(t)&=&\mbox{Tr}\,\langle 0|\psi^{\bf Q}_{\pm 1/2}(t)\bar\psi^{\bf Q}_{\pm 1/2}(0)|0\rangle\, .
\en 
It can be straightforwardly seen that, in the limit $t'\to+\infty$, $t\to-\infty$,
\eq\label{eq:tt1}
{\cal N}\frac{\mbox{Tr}\,(\tilde R_{1/2}(t',t))}{\tilde D_{1/2}(t'-t)}\,
\biggl(\frac{D^+_{\bf Q}(t')\tilde D_{1/2}(-t)\tilde D_{1/2}(t'-t)}
{\tilde D_{1/2}(t')D^+_{\bf Q}(-t)D^+_{\bf Q}(t'-t)}\biggr)^{1/2}&\to&
\langle 1/2| J^3(0)|1/2\rangle\, ,
\nonumber\\[2mm]
-{\cal N}\frac{\mbox{Tr}\,(R_{1/2}(t',t))}{D_{1/2}(t'-t)}\,
\biggl(\frac{D^-_{\bf Q}(t')D_{1/2}(-t)D_{1/2}(t'-t)}
{D_{1/2}(t')D^-_{\bf Q}(-t)D^-_{\bf Q}(t'-t)}\biggr)^{1/2}&\to&
\langle 1/2| J^+(0)|-1/2\rangle\, ,
\nonumber\\[2mm]
-{\cal N}\frac{\mbox{Tr}\,(R_{3/2}(t',t))}{D_{3/2}(t'-t)}\,
\biggl(\frac{D^+_{\bf Q}(t')D_{3/2}(-t)D_{3/2}(t'-t)}
{D_{3/2}(t')D^+_{\bf Q}(-t)D^+_{\bf Q}(t'-t)}\biggr)^{1/2}&\to&
\langle 3/2| J^+(0)|1/2\rangle\, ,
\en
where ${\cal N}=\sqrt{4E_\Delta E_N}$ 
and the Euclidean analogs of Eqs.~(\ref{eq:proj}) read (cf., e.g.,
 with Ref.~\cite{Alexandrou:2007dt}\footnote{Note the difference in sign and in a factor 2 with the third line of Eq.~(4) of Ref.~\cite{Alexandrou:2007dt}.})
\eq\label{eq:proj_e}
\langle 1/2|J^3(0)|1/2\rangle
&=&\frac{E_\Delta-Q^0}{E_\Delta}\,A\,G_C(t)
\nonumber\\[2mm]
\langle 1/2|J^+(0)|-1/2\rangle
&=&\sqrt{\frac{1}{2}}\,A
(G_M(t)-3G_E(t))\, ,
\nonumber\\[2mm]
\langle 3/2|J^+(0)|1/2\rangle
&=&\sqrt{\frac{3}{2}}A(G_M(t)+G_E(t))\, ,
\en
where $t=(E_\Delta-E_N)^2-{\bf Q}^2$ and the quantity $A$ is given
by Eq.~(\ref{eq:A}) with the replacement $E_R\to E_\Delta$. We would 
like to also mention that, in case of a stable $\Delta$, the above relations
hold up to the Lorentz-non-invariant terms exponentially suppressed in a
box of size $L$.

When the $\Delta$ becomes unstable, the interpretation of the above 
equations changes. The ratios given in Eq.~(\ref{eq:tt1}) can be still formed,
but the functions that are extracted from these ratios are the matrix elements
of the electromagnetic current calculated between a certain eigenstate of the
Hamiltonian in a finite volume 
(with the volume-dependent energy $E_\Delta$) 
and a one-nucleon state. The normalization constant ${\cal N}$ in these equations also changes. Namely, ${\cal N}=\sqrt{8w_{1\Delta}w_{2\Delta}E_N}$, where $w_{1\Delta}$ and $w_{2\Delta}$ are the energies of the nucleon and a pion in the CM system: 
$w_{1\Delta}=(E_\Delta^2+m_N^2-M_\pi^2)/(2E_\Delta)$,
$w_{2\Delta}=(E_\Delta^2-m_N^2+M_\pi^2)/(2E_\Delta)$ and 
$w_{1\Delta}+w_{2\Delta}=E_\Delta$. In the infinite-volume
limit, these matrix elements do not coincide with the resonance matrix elements
defined in the previous section. Rather, the energy of any fixed level tends
to the threshold value in this limit. Note also that, at a given energy,
there may exist several eigenstates of the Hamiltonian in the vicinity of the
resonance energy, and it is not clear, which of these matrix
elements should be identified with the resonance matrix element we are 
looking for.

It is evident that the situation closely resembles the determination of the 
resonance pole position from the lattice data by using the L\"uscher equation.
As it is well known, in order to achieve the goal, one has to perform an analytic continuation
into the complex energy plane. A detailed discussion of this procedure
can be found, e.g., in Ref.~\cite{Hoja2}. Below, we give a short description 
of the procedure. One first extracts the scattering
phase shift at different energies 
from the measured energy spectrum and fits the quantity 
$p^3\cot\delta(p)$ by a polynomial in $p^2$, assuming the effective 
range expansion (here, $p$ denotes the relative 3-momentum in the CM system). 
Then, one finds the poles of the $T$-matrix
in the complex plane by finding the zeros of a polynomial with known
coefficients. Below we shall prove the generalization of this procedure
to the case of the matrix elements. Namely, the matrix elements given in
Eq.~(\ref{eq:tt1}) should be measured at several energies (corresponding to the
measurement at several volumes), and the result should be fitted by some 
polynomial. Further, we shall show that, replacing $p^2$ by $p_R^2$ in this
polynomial, where $p_R$ denotes the value of the relative 3-momentum at 
the resonance pole, one obtains the
resonance form factors defined in the previous section (up to the 
corrections that are exponentially suppressed in large volumes). This 
is the main result of our work. Note also that this prescription is much 
simpler than the one for the elastic $\Delta$-form factor 
(see Ref.~\cite{Hoja2}) since, as we shall see below, no finite fixed points arise
in the case of the transition form factor.

As it is clear from the previous discussion, in order to perform
the fit, the matrix elements should be measured at several values of the
relative 3-momentum $p$. 
These matrix elements depend on two kinematic
variables: apart from $p$, there is the nucleon 3-momentum $|{\bf Q}|$
(alternatively, the variable $t$ which at the resonance takes the
complex value $t=(E_R-Q^0)^2-{\bf Q}^2$). Note also that fixing $|{\bf Q}|$
is equivalent to fixing $t$ because the (complex) resonance energy $E_R$, 
which is defined in the infinite volume, is also fixed.
On the lattice, it is more convenient to fix the real quantity $|{\bf Q}|$, and
we stick to this choice in the following.

According to the previous discussion, our goal is
 to find a way to ``scan'' the resonance region in the variable $p$,
leaving the other variable $|{\bf Q}|$ fixed. There is, however, a problem, if
one performs this scan by doing measurements at different volumes. Namely,
the momentum on the lattice along any axis is quantized, and the 
smallest nonzero momentum available is equal to $2\pi/L$, where $L$ denotes the box size. Consequently,
if $L$ is varied, the quantity $|{\bf Q}|$ will change along with $p$. 

We can propose at least two strategies that help to circumvent this problem:

\begin{figure}[t]
\begin{center}
\includegraphics[width=6.cm]{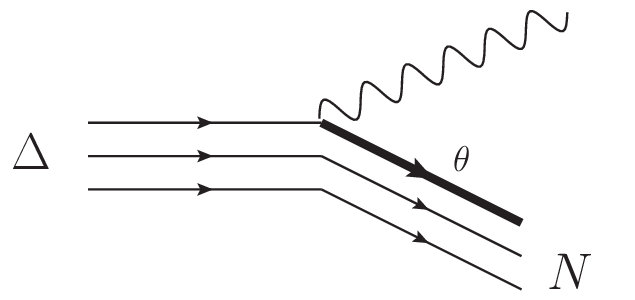}
\end{center}
\caption{Twisting a single quark in the nucleon.}
\label{fig:twisted_1q}
\end{figure}

\begin{enumerate}
\item
The use of asymmetric boxes. Consider the box with the geometry
$L\times L\times L'$ and direct ${\bf Q}$ along the third axis. The $\Delta$
is in the rest frame. Changing $L$ does not affect ${\bf Q}$ but affects $p$.

\item
Using (partially) twisted boundary conditions. One may apply the twisting to a single quark
in the nucleon, namely, the one that is attached to the photon 
(see Fig.~\ref{fig:twisted_1q}). This gives an additional momentum
to the nucleon along the third axis. The magnitude of this change is
$|{\bf Q}_\theta|=\theta/L$. On the other hand, changing the cubic box size,
we also change the magnitude of the nucleon momentum $|{\bf Q}|$. One may adjust
the value of the twisting angle $\theta$ so that the sum of these two effects
cancels and the nucleon momentum is kept fixed. It is important to stress that
the value of $\theta$ can be determined prior to the simulations since it
depends only on the box size. We also note that a similar technique of twisting
has been already applied in the past for the calculation of nucleon form 
factors~\cite{Schierholz-ptwist}.

\end{enumerate}

To summarize, on the lattice we have to measure the matrix elements given in 
Eq.~(\ref{eq:tt1}). These matrix elements depend on two kinematic variables:
the relative 3-momentum $p$ in the $\Delta$-channel and the 3-momentum ${\bf Q}$
of the nucleon (the three-momentum of the photon is $-{\bf Q}$). Using asymmetric boxes or twisted boundary conditions, we may
scan the resonance region in $p$ while keeping ${\bf Q}$ fixed. Let us now
discuss how to perform the analytic continuation into the 
complex plane and extract the resonance form factors with the use of
Eq.~(\ref{eq:proj_e}).

\section{Matrix elements in a finite volume}
\label{sec:diagrams}

\subsection{Two-point function}

In this section, we shall consider the resonance
matrix elements by using the technique of the non-relativistic effective field theory in a
finite volume. While doing so, we closely follow the path of
Ref.~\cite{Hoja2}, adapting the formulae given there, whenever necessary. To avoid problems, related to the mixing of the partial
waves, the $\Delta$-resonance is always considered in the CM frame. In this
case, there is no $S$- and $P$-wave mixing. Neglecting the (small) $P_{31}$
wave, the L\"uscher equation~\cite{Luescher-torus} for the $P_{33}$ wave is written as
follows:
\eq
p\cot\delta(p)+p\cot\phi(q)=0\, ,\quad\quad q=\frac{pL}{2\pi}\, ,
\en
where
\eq
p\cot\phi(q)=-\frac{2}{\sqrt{\pi}L}\,\biggl\{\hat Z_{00}(1;q^2)\pm\frac{1}{\sqrt{5}q^2}\,
\hat Z_{20}(1;q^2)\biggr\}\, ,
\en
and $\delta(p)$ is the $P_{33}$ phase shift in the infinite volume.
Further, $\hat Z_{lm}(1;q^2)$ denotes the L\"uscher zeta-function (for the
asymmetric boxes, in general), and the signs $+$ and $-$
are chosen for the irreps $G_1$  and $G_2$ of the little group corresponding to
${\bf d}=(0,0,1)$, respectively (see Ref.~\cite{Schierholz-group}). In
particular, for a symmetric box,  $\hat Z_{lm}(1;q^2)= Z_{lm}(1;q^2)$ and
$Z_{20}(1;q^2)=0$ in the CM frame (there is no mixing to the $P_{31}$ wave in
this case). For an asymmetric box with $L'=xL$,
\eq
\hat Z_{lm}(1;q^2)=\frac{1}{x}\sum_{\bf n\in\mathbb{Z}^3}\frac{{\cal Y}_{lm}({\bf r})}
{{\bf r}^2-q^2}\, ,\quad\quad r_{1,2}=n_{1,2}\, ,\quad r_3=\frac{1}{x}\,n_3\,
,\quad\quad \hat Z_{20}(1;q^2)\neq 0\, .
\en  
The matrix element of an operator ${\cal O}_i$
between the vacuum and an eigenstate of a
Hamiltonian $\langle 0|{\cal O}_i(0)|n\rangle$
 contains two-particle reducible diagrams describing initial-state
interactions, which are volume-dependent
(here, ${\cal O}_i$ stands for one of the operators
$\tilde {\cal O}_{1/2},{\cal O}_{1/2},{\cal O}_{3/2}$). 
This matrix element is proportional to $U_i$ even in case of an unstable
$\Delta$, where
\eq
U_{3/2}&=&\frac{1}{2}\,(1+\Sigma_3)\frac{1}{2}\,(1+\gamma_4)\frac{1}{\sqrt{2}}\,(u^1(P,3/2)-i\Sigma_3 u^2(P,3/2))\, ,
\nonumber\\[2mm]
U_{1/2}&=&\frac{1}{2}\,(1-\Sigma_3)\frac{1}{2}\,(1+\gamma_4)\frac{1}{\sqrt{2}}\,(u^1(P,1/2)+i\Sigma_3 u^2(P,1/2))\, ,
\nonumber\\[2mm]
\tilde{U}_{1/2}&=&\frac{1}{2}\,(1+\Sigma_3)\frac{1}{2}\,(1+\gamma_4)u^3(P,1/2)\, .
\en
This fact can be verified straightforwardly, since both the above matrix element
as well as $U_i$ are Dirac spinors with only one nonzero entry.

\begin{figure}[t]
\begin{center}
\includegraphics[width=8.cm]{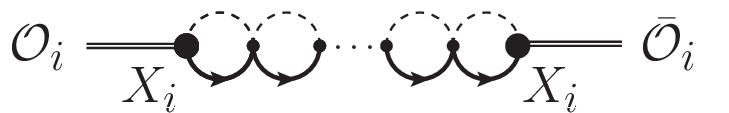}
\end{center}
\caption{Initial-state pion-nucleon interactions in the two-point function.
The quantity $X_i$ stands for the coupling of the operator ${\cal O}_i$ to 
the pion-nucleon pair in the intermediate state.  }
\label{fig:irr_twopoint}
\end{figure}

The calculation of the volume-dependent factor in the matrix element proceeds
by using the same technique as in the
derivation of Eq.~(46) of Ref.~\cite{Hoja2}. Namely, we calculate the two-point 
function of the operators ${\cal O}_i,\bar {\cal O}_i$ in the 
non-relativistic effective field theory below the inelastic threshold. 
According to the discussion above, 
the coupling of the operator ${\cal O}_i$ to the pion-nucleon state in the
effective theory is described by a local vertex $X_iU_i$, where 
the scalar function $X_i=X_i^{(0)}+X_i^{(1)}{\bf p}^2+\cdots$ contains the
terms with $0,2,\ldots$ derivatives. Here, ${\bf p}^2$ stands for the relative
momentum squared of the pion-nucleon pair in the CM system. Note that the 
$X_i$ contain only short-range physics and are the same in a finite and in the
infinite volume. Explicit values of $X_i^{(m)}$ are not important, because the
$X_i$ cancel in the final expressions.

Summing now up the pion-nucleon bubble diagrams as shown in Fig.~\ref{fig:irr_twopoint}, it is seen that the 
Euclidean two-point function takes the form
\eq\label{eq:twopoint}
\langle 0|{\cal O}_i(x_0)\bar {\cal O}_i(y_0)|0\rangle=
U_iX_i\biggl\{\int_{-\infty}^\infty\frac{dP_0}{2\pi}\,e^{iP_0(x_0-y_0)}
V(-iJ(P_0)-J^2(P_0)T(P_0))\biggr\}X_i\bar U_i\, ,
\en
where
\eq\label{eq:J0}
J(P_0)=\frac{1}{V}\,\sum_{\bf k}\frac{1}{2w_1({\bf k})2w_2({\bf k})}\,
\frac{1}{P_0-i(w_1({\bf k})+w_2({\bf k}))}\, ,
\en
$w_1({\bf k})=\sqrt{m_N^2+{\bf k}^2}\,, ~w_2({\bf k})=\sqrt{M_\pi^2+{\bf k}^2}$,
$V$ is the lattice volume and 
$T(P_0)$ denotes the pion-nucleon scattering amplitude in a finite
volume\footnote{Note that here we use a different normalization in the
  partial-wave expansion of the  $T$-matrix than in Ref.~\cite{Hoja2}.}
\eq\label{eq:T0}
T(P_0)=\frac{8\pi\sqrt{s}}{p\cot\delta(p)+p\cot\phi(q)}\, ,\quad\quad
s=-P_0^2\, ,
\en
and $p$ is the relative momentum in the CM frame, corresponding to the
total energy $\sqrt{s}$.

Next, we perform the integration over the variable $P_0$, using Cauchy's theorem.
It can be shown that only the poles of $T(P_0)$ contribute to this integral.
In the vicinity of the $n$-th pole, this function is given by
\eq
T(P_0)=\frac{8\pi E_n^2}{w_{1n}w_{2n}}\,\frac{\sin^2\delta(p_n)}
{\delta'(p_n)+\phi'(q_n)}\,\frac{1}{E_n+iP_0}+\cdots\, ,
\en  
where $E_n$ are the eigenenergies
in the box, $p_n$ is the corresponding relative 3-momentum, $q_n=p_nL/(2\pi)$,
and $w_{1n}=(E_n^2+m_N^2-M_\pi^2)/(2E_n)$,
$w_{2n}=(E_n^2-m_N^2+M_\pi^2)/(2E_n)$. 
The derivatives are taken with respect to the variable $p$, so that
$\phi'(q)=d\phi(q)/dp=(L/2\pi)d\phi(q)/dq$.

Performing the integration over $p_0$ and using the L\"uscher equation
\eq
\frac{1}{V}\sum_{\bf k}\frac{1}{2w_1({\bf k})2w_2({\bf k})}\,\frac{1}{w_1({\bf k})+w_2({\bf k})-E_n}=\frac{p_n\cot\delta(p_n)}{8\pi E_n}\, ,
\en
we finally get
\eq
\langle 0|{\cal O}_i(x_0)\bar {\cal O}_i(y_0)|0\rangle
=U_iX_i\biggl\{
V\sum_ne^{-E_n(x_0-y_0)}\frac{\cos^2\delta(p_n)}{\delta'(p_n)+\phi'(q_n)}\,
\frac{p_n^2}{8\pi w_{1n}w_{2n}}\biggr\}X_i\bar U_i\, .
\en
On the other hand,
\eq
\langle 0|{\cal O}_i(x_0)\bar {\cal O}_i(y_0)|0\rangle
=\sum_n\frac{e^{-E_n(x_0-y_0)}}{4w_{1n}w_{2n}}\,
\langle 0|{\cal O}_i(0)|n\rangle\langle n|\bar {\cal O}_i(0)|0\rangle\, .
\en
Comparing these two equations, we finally get
\eq\label{eq:2}
|\langle 0|{\cal O}_i(0)|n\rangle|=U_i\,X_i
V^{1/2}\biggl(\frac{\cos^2\delta(p_n)}{|\delta'(p_n)+\phi'(q_n)|}\,
\frac{p_n^2}{2\pi}\biggr)^{1/2}\, .
\en

\begin{figure}[t]
\begin{center}
\includegraphics[width=10.cm]{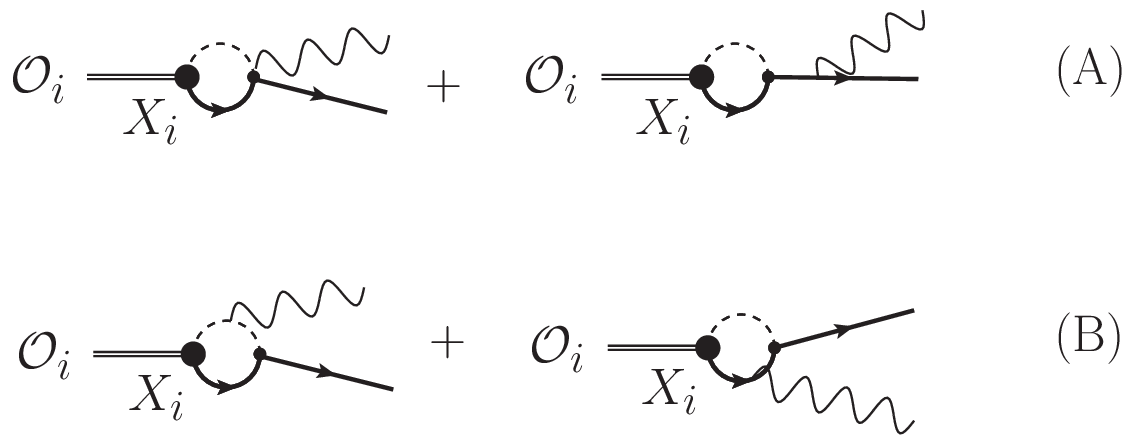}
\end{center}
\caption{Typical diagrams contributing to the $\Delta N\gamma^*$ 
transition form factor: (A) point vertex and emission of the photon from 
the external nucleon line; (B) emission of the photon from the internal lines.
As in Fig.~\ref{fig:irr_twopoint}, 
the quantity $X_i$ stands for the coupling of the operator ${\cal O}_i$ 
to the pion-nucleon pair in the intermediate state.}
\label{fig:irr_AB}
\end{figure}

\subsection{Three-point function}

\begin{figure}[t]
\begin{center}
\includegraphics[width=15.cm]{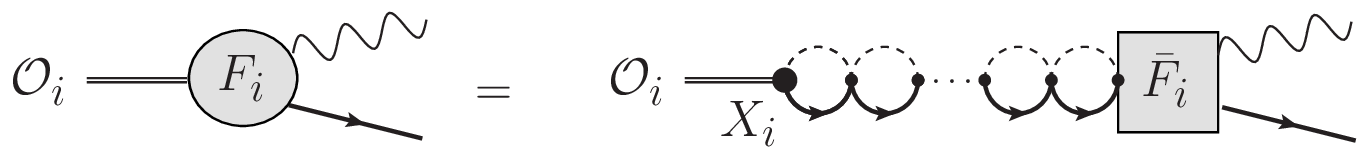}
\end{center}
\caption{Initial-state pion-nucleon interactions in the $\Delta N\gamma^*$ 
transition form factor. The quantity
$\bar F_i$ denotes the sum of all irreducible diagrams.}
\label{fig:irreducible}
\end{figure}

Next, let us consider the current matrix elements in a finite volume  
$F_i=F_i(p,|{\bf Q}|),~i=1,2,3$, which appear on the r.h.s. of
Eq.~(\ref{eq:tt1}). 
The derivation which is given below is essentially
similar to Eqs. (65)-(70) of Ref.~\cite{Hoja2}. We start from the calculation
of the three-point function, summing up the bubble diagrams
 in the non-relativistic effective theory. There are two types of diagrams,
which are shown in Fig.~\ref{fig:irr_AB} to which one has to add the
diagrams obtained by adding any number of pion loops to the initial 
states interaction (see
Fig.~\ref{fig:irreducible}). The self-energy insertions in the outgoing nucleon
line can be safely ignored since the nucleon is a stable particle (see discussion below) and hence such
diagrams lead to the exponentially suppressed contributions in a finite volume.
As a result, the matrix element is written as a sum of two contributions
\eq\label{eq:threepoint}
\langle 0|{\cal O}_i(x_0)J^a(0)|Q,\varepsilon\rangle
&=&U_iX_i V^{-1/2}\int_{-\infty}^\infty\frac{dP_0}{2\pi}\,e^{iP_0x_0}\frac{p\cot\delta(p)}{8\pi\sqrt{s}}
T(P_0)
\nonumber\\[2mm]
&\times&(-iJ(P_0)\bar F_i^{(A)}(p,|{\bf Q}|)+\hat F_i^{(B)}(p,|{\bf Q}|))\, .
\en
where  $F_i^{(A)}(p,|{\bf Q}|)$ does not depend on $L$ 
(up to the exponentially suppressed contributions)
and
the indices $a,\varepsilon$ have been suppressed for brevity.

The diagrams of the type (B) can be potentially dangerous. Indeed in the case of the elastic form factor, such diagrams lead to the so-called finite fixed points, see Ref.\cite{Hoja2}. However, the fact that one of the external particles (the nucleon) is stable simplifies matters considerably. As shown in section~\ref{sec:analytic} (see Eq.~(\ref{eq:I1})), in that case the quantity $\hat F_i^{(B)}(p,|{\bf Q}|))$ can be written as a product of two factors:
\eq
\hat F_i^{(B)}(p,|{\bf Q}|))=-iJ(P_0)\bar F_i^{(B)}(p,|{\bf Q}|))\, ,
\label{eq:B}
\en
where $\bar F_i^{(B)}(p,|{\bf Q}|))$, again, does  not depend on $L$ 
up to the exponentially suppressed contributions. 
Physically, this corresponds to the Taylor expansion of the pion
propagator attached to the nucleon in diagram  Fig.~\ref{fig:transition_culprit}:
this propagator shrinks to a point (a similar discussion holds when the photon is attached to the nucleon line) and the contributions of diagrams (B)
to the matrix element are equivalent to the one of diagrams (A).

Consequently,
we can rewrite Eq.~(\ref{eq:threepoint})
\eq\label{eq:threepoint1}
\langle 0|{\cal O}_i(x_0)J^a(0)|Q,\varepsilon\rangle
=U_iX_iV^{-1/2}\int_{-\infty}^\infty\frac{dP_0}{2\pi}\,e^{iP_0x_0}(-iJ(P_0))\frac{p\cot\delta(p)}{8\pi\sqrt{s}}
T(P_0)
\bar F_i(p,|{\bf Q}|)\, ,
\en
where $\bar F_i(p,|{\bf Q}|)=\bar F^{(A)}_i(p,|{\bf Q}|)+\bar F^{(B)}_i(p,|{\bf Q}|)$ denotes the full irreducible amplitude for the
$\pi N\to\gamma^*N$ transition.

Performing now the Cauchy integral over $P_0$ and comparing to the spectral
representation of the three-point function, we get
\eq\label{eq:3}
\langle 0|{\cal O}_i(0)|n\rangle \frac{1}{4w_{1n}w_{2n}}\,F_i(p_n,|{\bf Q}|)
=X_i\,U_iV^{-1/2}\,
\frac{\cos^2\delta(p_n)}{\delta'(p_n)+\phi'(q_n)}\,
\frac{p_n^2}{8\pi w_{1n}w_{2n}}\,\bar F_i(p_n,|{\bf Q}|)\, .
\en
From Eqs.~(\ref{eq:2}) and Eq.~(\ref{eq:3}) one obtains
\eq\label{eq:barFF}
|\bar F_i(p_n,|{\bf Q}|)|
=V\biggl(\frac{\cos^2\delta(p_n)}{|\delta'(p_n)+\phi'(q_n)|}
\frac{p_n^2}{2\pi} \biggr)^{-1/2}|F_i(p_n,|{\bf Q}|)|\, .
\en

Before we proceed further, an important remark is in order. As already discussed
in section~\ref{sec:infinite}, the form factor can be defined either on the
real energy axis, or at the resonance pole (although, strictly speaking, only
the latter definition is a rigorous one, the former is process-dependent).
Below we shall provide the formulae which enable one to ``translate'' lattice
data into one of these definitions.

\subsubsection{Real energy axis}
On the {\em real} energy axis, the infinite-volume matrix
element, corresponding to the scattering process $\pi N\to \gamma^*N$ at low energies, is given by a geometric series of the pion-nucleon bubbles in the 
final state. This series sums up into the following expression:
\eq\label{eq:Watson}
{\cal A}_i(p,|{\bf Q}|)=\frac{p\cot\delta(p)}{p\cot\delta(p)-ip}\,\bar F_i(p,|{\bf Q}|)
=e^{i\delta(p)}\cos\delta(p)\,\bar F_i(p,|{\bf Q}|)\, .
\en
Here, we have assumed that
 the quantity $\bar F_i(p,|{\bf Q}|)$ has a smooth infinite-volume
limit (see the proof below). The amplitudes ${\cal A}_i$ are proportional to 
the linear combinations of the so-called transverse magnetic ($M$), transverse 
electric ($E$) and scalar ($S$) multipoles. The latter appear in the partial 
wave decomposition of the $\pi N\to \gamma^*N$ scattering amplitude (see  
Appendix~\ref{app:photoproduction} for the details) and can be extracted from the analysis 
of the 
experimental data. The multipoles contain information on the resonance states\footnote{We would like to mention here that in Ref.~\cite{Meyer}
the calculation of the deuteron photodisintegration amplitude in a finite 
volume was addressed by using a slightly different technique.}.

Note that Eq.~(\ref{eq:Watson}) is nothing but Watson's theorem, which indeed holds near the $\Delta$-resonance in the elastic region. At a first glance, 
${\cal A}_i(p,|{\bf Q}|)$ vanishes when $\delta=90^{\sf o}$. 
In order to show that this is not the case, we 
rewrite the Eq.~(\ref{eq:Watson}) as follows:
\eq\label{eq:Watson1}
{\cal A}_i(p,|{\bf Q}|)=\frac{e^{i\delta(p)}}{p^3}\sin\delta(p)\, p^3\cot\delta(p)\,\bar F_i(p,|{\bf Q}|)\, ,
\en
Assuming that the effective-range expansion holds in the resonance region, one
may write
\eq\label{eq:effrange}
p^3\cot\delta(p)\doteq h(p^2)=-\frac{1}{a}+\frac{1}{2}\,rp^2+\cdots\, ,
\en
where $a$ is the $P$-wave scattering volume and $r$ is the effective range.
The function $h(p^2)$ should have a zero at $p^2=p_A^2$, where the scattering
phase passes through $90^{\sf o}$.
It is easy to get convinced that the quantity  $\bar F_i(p,|{\bf Q}|)$ should
have a pole exactly at the same value of $p^2$. This pole corresponds to the
exchange of the bare $\Delta$ in the $s$-channel (since 
our effective non-relativistic Lagrangian does not include the explicit 
$\Delta$, the pole
will manifest itself in the divergence on the perturbative series at
$p^2=p_A^2$). Consequently, not the quantity $\bar F_i(p,|{\bf Q}|)$ alone,
but the product $p^3\cot\delta(p)\bar F_i(p,|{\bf Q}|)$ is a low-energy
polynomial that can be safely expanded in the resonance region and that,
in general, does not vanish at $p^2=p_A^2$. It follows from this that the
multipoles ${\cal A}_i$, defined by Eq.~(\ref{eq:Watson1}), take
finite values at the resonance.

Finally, combining the above equations, we arrive at an analogue of the
L\"uscher-Lellouch equation for the photoproduction amplitude in the elastic 
region
\eq\label{eq:Main}
{\cal A}_i(p_n,|{\bf Q}|)=e^{i\delta(p_n)}\,
V\biggl(\frac{1}{|\delta'(p_n)+\phi'(q_n)|}
\frac{p_n^2}{2\pi}\biggr)^{-1/2}|F_i(p_n,|{\bf Q}|)|,
\en
The Eq.~(\ref{eq:Main}) comprises one of the main results of the present 
article. It allows to extract the multipole amplitudes from lattice data.

In order to obtain the $\Delta N \gamma^*$ matrix elements $F^A_i$, defined
on the real axis,  
one parameterizes the imaginary parts of the multipoles 
through the matrix element of the electromagnetic current between $N$ and 
$\Delta$ states (see, e.g., Ref.~\cite{Aznauryan}). Then, 
in the narrow width approximation,
for the amplitudes ${\cal A}_i(p,|{\bf Q}|)$ we get:
\eq\label{eq:Aznauryan}
|{\rm Im}\,{\cal A}_i(p_A,|{\bf Q}|)|=\sqrt{\frac{8\pi}{p_A\Gamma}}|F_i^A(p_A,|{\bf Q}|)|,
\en
where all quantities are {\it real} and taken at the Breit-Wigner pole $p=p_A$, the $\Gamma$ is the total width of the $\Delta$-resonance. 

\subsubsection{Complex energy plane}
\label{sec:analytic}

Next, we consider the extraction of the form factor at the resonance
pole. This implies the analytic continuation of the above result into
the complex $p$-plane. In order to do this, let us first consider the two-point
function in the infinite volume
\eq
\langle 0|O_i(x)\bar O_i(y)|0\rangle=
\int\frac{d^4P}{(2\pi)^4}\, e^{iP(x-y)} D_i(P^2)\, ,
\en
where, in the CM system $P_\mu=(P_0,{\bf 0})$ 
the quantity  $D_i(P^2)$ takes the form
 (cf.  Eq.~(\ref{eq:twopoint}))
\eq
D_i(P^2)=U_iX_i(-iJ_\infty(P_0)-J_\infty^2(P_0)T_\infty(P_0))X_i\bar U_i\, .
\en
Here, the quantities $J_\infty(P_0)$ and $T_\infty(P_0)$ denote the infinite-volume
counterparts of the quantities defined by Eqs.~(\ref{eq:J0}) and (\ref{eq:T0}).
In the Minkowski space, with $P_0=i\sqrt{s}$, these quantities are given by
\eq
J_\infty(P_0)=-\frac{p}{8\pi \sqrt{s}}\, ,\quad\quad
T_\infty(P_0)=\frac{8\pi \sqrt{s}}{p\cot\delta(p)-ip}\, .
\en
These expressions are valid on the first Riemann sheet. On the second sheet,
the relative momentum $p$ changes  sign.

Suppose now that the scattering amplitude $T_\infty(P_0)$ has a pole 
at $s=s_R$ on the second Riemann sheet.
Writing down the effective-range expansion in a form of Eq.~(\ref{eq:effrange}),
one first finds the pole position in the complex plane from the equation
\eq
-\frac{1}{a}+\frac{1}{2}\,rp_R^2+\cdots=-ip_R^3\, .
\en
Further, in the vicinity of the pole $p=p_R$ we get
\eq
p^2(p\cot\delta(p)+ip)=(s-s_R)(2p_Rh'(p_R^2)+3ip_R^2)\frac{w_{1R}w_{2R}}{2p_Rs_R}
+O((s-s_R)^2)\,
\en
where $w_{1R}=\sqrt{m_N^2+p_R^2}$, $w_{2R}=\sqrt{M_\pi^2+p_R^2}$,
$s_R=E_R^2=(w_{1R}+w_{2R})^2$ and the derivative of $h$ is taken with respect to 
the variable $p^2$. 
Using this expansion, one may obtain the value of the
wave function renormalization constant at the pole:
\eq
D_i(s)&\to&U_iX_i\, \frac{Z_R}{s_R-s} X_i\bar U_i+
\mbox{regular terms at $s\to s_R$}\, ,\quad\quad
\nonumber\\[2mm]
Z_R&=&\biggl(\frac{p_R}{8\pi E_R}\biggr)^2
\biggl(\frac{16\pi p_R^3 E_R^3}{w_{1R}w_{2R}(2p_Rh'(p_R^2)+3ip_R^2)}\biggr)\, .
\en
The three-point function in the infinite volume is given by
\eq\label{eq:infty3}
\langle 0|O_i(x)J^a(0)|Q,\varepsilon\rangle
=U_iX_i\int\frac{d^4P}{(2\pi)^4}\,e^{iPx}(-iJ_\infty(P_0))\frac{p\cot\delta(p)}{8\pi\sqrt{s}}\, T_\infty(P_0)\bar F_i(p,|{\bf Q}|)\, ,
\en
where $\bar F_i(p,|{\bf Q}|)$ is the same as in Eq.~(\ref{eq:threepoint1}).

Next, we assume that the two-particle irreducible part 
$\bar F_i(p_n,|{\bf Q}|)$ can be analytically continued into the complex plane. 
Separating the pole contribution in the three-point function,
we get our final expression for the resonance matrix element $F^R_i$, evaluated
at the pole
\eq\label{eq:formfactor}
F_i^R(p_R,|{\bf Q}|)=Z_R^{1/2}\,\bar F_i(p_R,|{\bf Q}|)\, .
\en
As a test of our final formula, let us consider the case when the resonance is infinitely narrow. Then, the pole tends to the real axis, and 
$E_R\to E_n,~p_R\to p_n$. Still, we assume that the (real) energy $E_R$
is above the two-particle threshold.

First, we  can express the amplitudes ${\cal A}_{i}$ through $F_i^R$:
\eq
{\cal A}_{i}(p_R,|{\bf Q}|)=Z_R^{-1/2}F_i^R(p_R,|{\bf Q}|).
\en
Further, since $h(p^2)=p^3\cot\delta(p)$, at the  resonance we have
\eq
2p_Rh'(p_R^2)+3ip_R^2=-\frac{p_R^3\delta'(p_R)}{\sin^2\delta(p_R)}\, .
\en
Moreover, in the vicinity of an infinitely narrow resonance,
 the derivative of the phase shift behaves as
\eq\label{eq:pshift}
\delta'(p_R)=\frac{2}{\Gamma}\frac{p_R E_R}{w_{1R}w_{2R}}\, ,
\en
where $\Gamma$ is the Breit-Wigner width of the resonance.
The renormalization constant $Z_R$ becomes
\eq
Z_R=-\frac{p_R\Gamma}{8\pi},
\en
and we finally obtain
\eq
|{\rm Im}\,{\cal A}_{i}(p_R,|{\bf Q}|)=\sqrt{\frac{8\pi}{p_R\Gamma}}|F_i^R(p_R,|{\bf Q}|)|,
\en
where $p_R\to p_A$. This formula coincides exactly with 
Eq.~(\ref{eq:Aznauryan}).

Further, in the limit considered, the derivative of the phase shift 
explodes (see Eq.\,(\ref{eq:pshift})) whereas the quantity 
$\phi'(q_n)$ stays finite.
Consequently, one may neglect $\phi'(q_n)$ in all formulae.
Expressing the quantity $\bar F_i(p_R,|{\bf Q}|)$ through $F_i(p_R,|{\bf Q}|)$
by using Eq.~(\ref{eq:barFF}) and substituting in Eq.~(\ref{eq:formfactor}),
we arrive at a fairly simple result in this limit:
\eq
F_i^R(p_n,|{\bf Q}|)=V\biggl(\frac{E_n}{2w_{1n}w_{2n}}\biggr)^{1/2}
F_i(p_n,|{\bf Q}|)\, .
\en
Note that the factor in front of $F_i$  on the r.h.s. of the above equation exactly accounts for
the difference in the normalization of the one- and two-particle states in
a finite volume (we remind the reader that the energy $E_R$ lies above 
threshold, so that the resonance still decays into the two-particle state, 
albeit with an infinitesimally small rate).

\subsection{Analytic continuation of the loop diagram}
\begin{figure}[t]
\begin{center}
\includegraphics[width=5.cm]{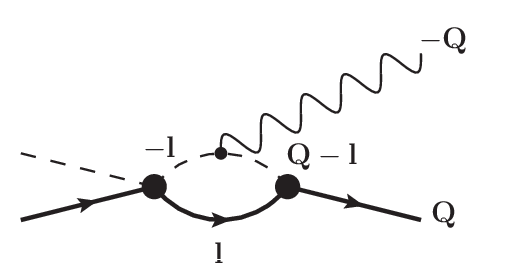}
\end{center}
\caption{A potentially dangerous diagram which involves the irreducible vertex $\bar F_i$.
A diagram where the photon is attached to the nucleon line, can be treated similarly. The case when the photon is attached to a local $\pi N\to\gamma^*N$ vertex is trivial, because, obviously, the corresponding diagram is a low-energy polynomial.}
\label{fig:transition_culprit}
\end{figure}

Finally, we want to demonstrate that the analytic continuation 
of the quantity  $\bar F_i(p_n,|{\bf Q}|)$ into the complex plane is possible.
In case of the elastic form factor, this procedure has led to serious
difficulties due the the existence of the so-called finite fixed points~\cite{Hoja2}.
However, such difficulties do not arise in case of the transition form factors,
as can be easily seen by considering the potentially dangerous triangle
diagram where the photon is attached to the pion line (see Fig.~\ref{fig:transition_culprit}). For simplicity, we neglect all numerators, which are  low-energy polynomials. In the rest-frame of the
$\Delta$-resonance, the diagram shown in Fig.~\ref{fig:transition_culprit} 
is equal to 
\eq\label{eq:I}
I=\frac{1}{V}\sum_{\bf l}\frac{1}{8w_1({\bf l})w_2(-{\bf l})w_2({\bf Q}-{\bf l})}\,
\frac{1}{(w_1({\bf l})+w_2(-{\bf l})-E_n)(w_1({\bf l})+w_2({\bf Q}-{\bf
    l})-Q^0)}\, .
\en
This diagram can be simplified by using the following algebraic identity
(see Ref.~\cite{Hoja1})
\eq
\frac{1}{4w_1({\bf l})w_2(-{\bf l})}\,
\frac{1}{(w_1({\bf l})+w_2(-{\bf l})-E_n)}
=\frac{1}{2E_n({\bf l}^2-p_n^2)}+\mbox{non-singular terms}\, .
\en

Since the second denominator in Eq.~(\ref{eq:I}) is non-singular, up to the
exponentially suppressed terms we get
\eq\label{eq:I1}
I=\frac{1}{V}\sum_{\bf l}\frac{1}{2E_n({\bf l}^2-p_n^2)}\,
\frac{1}{2}\int_{-1}^{1} dy \frac{1}{2\hat w_2
(\hat w_1+\hat w_2-Q^0)}\, ,
\en
where
\eq
\hat w_1=\sqrt{m_N^2+p_n^2}\, ,
\quad\quad
\hat w_2=\sqrt{M_\pi^2+p_n^2+{\bf Q}^2-2|{\bf Q}|p_ny}\, .
\en
Since the first factor on the r.h.s. of Eq.~(\ref{eq:I1}) can be replaced
by $p_n\cot\delta(p_n)$ from the L\"uscher equation, and the remaining
integral over $y$ is a low-energy polynomial, we see that the quantity $p^2I$ is
a low-energy polynomial in $p^2$ as well. No finite fixed points arise and the
analytic continuation can be performed without any problem. Moreover, there
exist only exponentially suppressed corrections to the infinite-volume limit.
Finally, it is now easy to check that the quantity 
$\hat F_i^{(B)}(p,|{\bf Q}|)$ in  Eq.~(\ref{eq:threepoint}) can be decomposed as in Eq.~(\ref{eq:B}), with the irreducible vertex 
$\bar F_i^{(B)}(p,|{\bf Q}|)$ containing only
exponentially suppressed finite-volume corrections. 
The same method can be used for the calculation of the three-point function in the infinite volume, see Eq.~(\ref{eq:infty3}). The sum over the momentum
${\bf l}$ in Eq.~(\ref{eq:I1}) is replaced by the integral and gives a pion-nucleon loop, whereas the remainder is again identified with the irreducible vertex
$\bar F_i^{(B)}(p,|{\bf Q}|)$.

\section{A prescription for the measurement of the transition form factors}
\label{sec:prescription}

This section contains a short summary of all our findings.
We give a prescription for calculating the $\Delta N\gamma^*$ transition
form factors on the lattice, in the rest-frame of the $\Delta$-resonance:

\begin{enumerate}

\item
The matrix elements $F_i=F_i(p,|{\bf Q}|)$ in the right-hand side of
Eq.~(\ref{eq:tt1}) are functions of the kinematic variables $p$ and $|{\bf
  Q}|$. Measure these matrix elements at different values of the variable $p$
in the resonance region, keeping the other variable fixed, as explained in
section~\ref{sec:setting}. The scattering phase 
should be measured at the same values of $p$.

\item
The multipoles for the pion photoproduction are given by
\eq
{\cal A}_i(p_n,|{\bf Q}|)=e^{i\delta(p_n)}\, 
V\biggl(\frac{p_n^2}{2\pi|\delta'(p_n)+\phi'(q_n)|}\biggr)^{-1/2}
|F_i(p_n,|{\bf Q}|)|\, .
\en

\item
The resonance matrix elements, defined at real energies,
are proportional to the imaginary part of the multipoles at
$p=p_A$, where the phase shift passes through $90^{\sf o}$. In the narrow
 width approximation one has
\eq
|{\rm Im}\,{\cal A}_i(p_A,|{\bf Q}|)|
=\sqrt{\frac{8\pi}{p_A\Gamma}}|F_i^A(p_A,|{\bf Q}|)|.
\en

\item
In order to extract the matrix element at the resonance pole,
we first multiply each $F_i$ by the pertinent L\"uscher-Lellouch factor
\eq
\bar F_i(p_n,|{\bf Q}|)
=V\biggl(\frac{\cos^2\delta(p_n)}{|\delta'(p_n)+\phi'(q_n)|}
\frac{p_n^2}{2\pi}\biggr)^{-1/2}F_i(p_n,|{\bf Q}|)\, .
\en

\item
Further, we
fit the functions $p^3\cot\delta(p)\,\bar F_i(p,|{\bf Q}|)$ by the effective-range formula
\eq
p^3\cot\delta(p)\,\bar F_i(p,|{\bf Q}|)=A_i(|{\bf Q}|)+p^2B_i(|{\bf Q}|)+\cdots\, .
\en

\item
Finally, we evaluate the resonance matrix elements by substitution
\eq
F_i^R(p_R,|{\bf Q}|)=i\,p_R^{-3}\,Z_R^{1/2}(A_i(|{\bf Q}|)+p_R^2B_i(|{\bf Q}|)+\cdots)\, .
\en
The quantities $p_R$ and $Z_R$ should be evaluated separately from the
measured phase shifts. Note that the form factors are related to the resonance matrix elements via the
formulae given in Eq.~(\ref{eq:proj_e}). 
The kinematic factors in front of the form factors
are low-energy polynomials. 

We emphasize again that, from the two definitions of the resonance
matrix elements, given above, only the one which implies the analytic 
continuation to the resonance pole, yields the result which is 
process-independent.

\end{enumerate}

\section{Conclusions}
\label{sec:concl}

\begin{itemize}

\item[i)]
In this paper, we have formulated an explicit prescription for the measurement
of  the $\Delta N\gamma^*$ transition form factors on the lattice. The $\Delta$
is 
considered as a resonance, not as a stable particle. The spins of all
particles are included, and three different scalar form factors are
projected out. 
\item[ii)]
The framework is based on the use of the non-relativistic effective field
theory in a finite volume. This is in accordance with the assumption of
validity of the effective range expansion in the vicinity of a resonance that
is used for performing the analytic continuation into the complex plane. If
this assumption proves to be very restrictive, our approach can be easily
adapted for the use of the alternative techniques (e.g., expanding the
amplitude in the vicinity of some point near the resonance energy, rather than
expanding around threshold).
\item[iii)]
The extraction of the elastic resonance form factors from data is a rather
subtle procedure due to the
presence of the so-called finite fixed points.
In case of the transition form factors, considered in the present paper, 
 the method is straightforward. The complexity of the extraction is
similar to the one of
determining the energy and width of the
$\Delta$-resonance. For this reason, we believe that the lattice study of the
transition form factors may become feasible in a foreseeable future.

\item[iv)]
The extraction of the form factors have been carried out in the rest-frame of
the $\Delta$-resonance. There are no serious obstacles to carrying out the
same procedure in the moving frames as well, except the mixing between $S$-
and $P$-waves which takes place, if the $\Delta$ is not at the rest.

\end{itemize}

{\em Acknowledgments:} 
The authors thank J. Gegelia, Ch. Lang, H. Meyer, S. Prelovsek, A. Sarantsev, S. Sharpe and G. Schierholz
for interesting discussions. 
This work is partly supported by the EU
Integrated Infrastructure Initiative HadronPhysics3 Project  under Grant
Agreement no. 283286. We also acknowledge the support by the DFG (CRC 16,
``Subnuclear Structure of Matter''), by
the Shota Rustaveli National Science Foundation
(Project DI/13/02) and by the Bonn-Cologne Graduate School of Physics and
Astronomy. This research is supported in part by Volkswagenstiftung
under contract no. 86260.

\renewcommand{\thefigure}{\thesection.\arabic{figure}}
\renewcommand{\thetable}{\thesection.\arabic{table}}
\renewcommand{\theequation}{\thesection.\arabic{equation}}

\appendix

\setcounter{equation}{0}
\setcounter{figure}{0}
\setcounter{table}{0}

\section{Photoproduction amplitudes}
\label{app:photoproduction}

In this appendix, we shall establish the connection between our 
photoproduction amplitude, defined in non-relativistic effective field 
theory with the relativistic amplitude. Note that in the present paper we deal 
with the $P$-wave in the $\gamma^*p\rightarrow\pi^0p$ channel with 
the total isospin $I=3/2$ and total spin $J=3/2$. Below, we give the 
expressions, using the relativistic normalization of the Dirac 
spinors~\cite{IZ}.

The relativistic photoproduction amplitude can be written in the rest frame of the pion-nucleon system as
\eq\label{eq:tmatrix}
T=8\pi E\, \chi^{\dagger}(2){\cal F}\chi(1),
\en
where $E$ is the total energy of the $\pi N$ system. The Pauli spinors $\chi(1)$, $\chi(2)$ carry the information on the spin states of the nucleons.

The matrix ${\cal F}$ has a decomposition (see, e.g., \cite{Drechsel})
\eq \label{eq:Drechsel}
{\cal F} & = & i\tilde{\bs\sigma}\cdot{\bs\epsilon}F_{1}
+({\bs\sigma}\cdot\hat{\bs q})({\bs\epsilon}\cdot({\bs\sigma}\times\hat{\bs k}))F_{2}
+i(\tilde {\bs q}\cdot{\bs\epsilon})({\bs\sigma}\cdot\hat{\bs k})F_{3} 
+i(\tilde {\bs q}\cdot{\bs\epsilon})({\bs\sigma}\cdot \hat {\bs q})F_{4}\nonumber\\
& &  +i(\hat {\bs k}\cdot{\bs\epsilon})({\bs\sigma}\cdot\hat{\bs k})F_{5}
+i(\hat {\bs k}\cdot{\bs\epsilon})({\bs\sigma}\cdot\hat{\bs q})F_{6}-\epsilon_0\left[i({\bs\sigma}\cdot 
\hat {\bs q})F_{7}
+i({\bs\sigma}\cdot \hat {\bs k})F_{8}\right]\,,
\en
where $\epsilon^\mu=(\epsilon_0,\,\bs\epsilon)$ is the photon polarization vector, $\hat {\bs k}={\bs k}/| {\bs k}| $ 
and $\hat {\bs q}={\bs q}/| {\bs q}| $ are the unit 
vectors for the photon and pion momenta  
respectively, and $\tilde{\bs a}={\bs a}-({\bs a}\cdot\hat{\bs k})
\hat{\bs k}$ is a vector with purely transverse components.     
The eight amplitudes $F_1,...,F_8$ are functions of three 
independent variables, e.g. the total energy $E$, the pion angle 
$\theta$, and the four-momentum squared of the virtual photon, 
$Q^2={\bs k}^2-\omega^2 >0$. 

Current conservation additionally implies that 
\eq
 |{\bs k}| F_{5}=\omega F_{8},\quad
|{\bs k}| F_{6}=\omega F_{7}.
\en
Further, the six independent amplitudes $F_1,...,F_6$ have a multipole decomposition
\begin{eqnarray} \label{eq:multipoles}
F_{1} & = & \sum_{l\geq0}\{(lM_{l+}+E_{l+})P_{l+1}^{\prime}
+[(l+1)M_{l-}+E_{l-}]P_{l-1}^{\prime}\}, \nonumber\\
F_{2} & = & \sum_{l\geq1}[(l+1)M_{l+}+lM_{l-}]P_{l}^{\prime}, \nonumber\\
F_{3} & = & \sum_{l\geq1}[(E_{l+}-M_{l+})P_{l+1}^{\prime\prime}
+((E_{l-}+M_{l-})P_{l-1}^{\prime\prime}],   \nonumber\\
F_{4} & = & \sum_{l\geq2}(M_{l+}-E_{l+}-M_{l-}-E_{l-})P_{l}^{\prime\prime}, \\
F_{5} & = & \sum_{l\geq0}[(l+1)L_{1+}P_{l+1}^{\prime}
-lL_{l-}P_{l-1}^{\prime}],  \nonumber\\
F_{6} & = & \sum_{l\geq1}[lL_{1-}-(l+1)L_{l+}]P_{l}^{\prime}\,,\nonumber
\end{eqnarray}
where $l$ is a pion angular momentum and the sign $\pm$ refers to the total spin $J=l\pm 1/2$. The $P_{l}^{\prime}$ are the derivatives of the Legendre polynomials $P_l=P_l(\cos\theta)$.
Note that in the literature the longitudinal transitions are often 
described by $S_{l\pm}$ multipoles, related to the $L_{l\pm}$ by 
\eq
S_{l\pm}=| {\bs k}| L_{l\pm}/\omega.
\en
In case of scattering
in the channel with the quantum numbers of the $\Delta$-resonance,
we  retain only $\ell\pm=1+$ partial wave and choose the
momentum ${\hat {\bs k}}$ along the third axis.
The pertinent scattering amplitudes then take the form
\begin{eqnarray}\nonumber\label{eq:tmelements}
\tilde{T}_{1/2}&=&\sqrt{4\pi}\left[\sqrt{\frac{1}{3}}\chi^\dagger_{-1/2}Y_{11}(\hat{\bs q})+\sqrt{\frac{2}{3}}\chi^\dagger_{1/2}Y_{10}(\hat{\bs q})\right]\chi_{1/2}\,\tilde{{\cal A}}_{1/2},\\\nonumber
T_{1/2}&=&\sqrt{4\pi}\left[\sqrt{\frac{1}{3}}\chi^\dagger_{-1/2}Y_{11}(\hat{\bs q})+\sqrt{\frac{2}{3}}\chi^\dagger_{1/2}Y_{10}(\hat{\bs q})\right]\chi_{-1/2}\,{\cal A}_{1/2},\\
T_{3/2}&=&\sqrt{4\pi}\left[\chi^\dagger_{1/2} Y_{11}(\hat{\bs q})\right]\chi_{1/2}\,{\cal A}_{3/2}\, .
\end{eqnarray}
Further, in order to relate the amplitudes ${\cal A}_{i}$ to the 
multipoles $M,E,S$, we 
make the following choice of the polarization vectors
\begin{eqnarray}\nonumber\label{eq:cases}
i=1:&\quad&\epsilon_0=\frac{|\bs k|}{Q},
\quad {\bs\epsilon}=\frac{1}{Q}(0,0,\omega)\,,
\\[2mm]\nonumber
i=2:&\quad&\epsilon_0=0,\quad {\bs\epsilon}=\frac{1}{\sqrt{2}}(1,i,0)\, ,
\\[2mm]
i=3:&\quad&\epsilon_0=0,\quad {\bs\epsilon}=\frac{1}{\sqrt{2}}(1,i,0)\, .
\end{eqnarray}
Below, we shall demonstrate the procedure explicitly for the choice $i=1$.
Retaining only the $\ell\pm=1+$ partial wave in 
Eq. (\ref{eq:multipoles}) and substituting the explicit polarization vectors
from Eq.~(\ref{eq:cases}) into Eq.~(\ref{eq:Drechsel}), one obtains:
\eq
{\cal F}=\frac{Q}{|{\bs k}|}\left[-6i({\bs\sigma}\cdot 
\hat {\bs k})\cos\theta
+2i({\bs\sigma}\cdot \hat {\bs q})\right]S_{1+}
\en
The expression $\chi^{\dagger}(2)\,({\bs\sigma}\cdot 
\hat {\bs k})\cos\theta\,\chi(1)$ can be brought into the form
\eq
\chi^{\dagger}(2)\,({\bs\sigma}\cdot 
\hat {\bs k})\cos\theta\,\chi(1)&=&\frac{\sqrt{2}}{3}\sqrt{4\pi}\left[\sqrt{\frac{1}{3}}\chi^\dagger_{-1/2}Y_{11}(\hat{\bs q})+\sqrt{\frac{2}{3}}\chi^\dagger_{1/2}Y_{10}(\hat{\bs q})\right]\chi(1)
\nonumber\\[2mm]
&-&\frac{1}{3}\sqrt{4\pi}\left[\sqrt{\frac{2}{3}}\chi^\dagger_{-1/2}Y_{11}(\hat{\bs q})-\sqrt{\frac{1}{3}}\chi^\dagger_{1/2}Y_{10}(\hat{\bs q})\right]\chi(1)\, .
\en
On the other hand, since
\eq
\chi^{\dagger}(2)\,({\bs\sigma}\cdot \hat {\bs q})\,\chi(1)
=\sqrt{4\pi}\left[\sqrt{\frac{2}{3}}\chi^\dagger_{1/2}Y_{10}(\hat{\bs q})-\sqrt{\frac{1}{3}}\chi^\dagger_{-1/2}Y_{11}(\hat{\bs q})\right]\chi(1)\, ,
\en
the quantity $\chi^{\dagger}(2)\,({\bs\sigma}\cdot \hat {\bs q})\,\chi(1)$ gives a $J=1/2$ contribution only. Consequently, the $J=3/2$ partial wave  contribution to the relativistic amplitude of Eq. (\ref{eq:tmatrix}) is
\eq
\tilde{T}_{1/2}=\left(-16\pi iE\sqrt{2}\frac{Q}{|{\bs k}|}S_{1+}\right)\sqrt{4\pi}\left[\sqrt{\frac{1}{3}}\chi^\dagger_{-1/2}Y_{11}(\hat{\bs q})+\sqrt{\frac{2}{3}}\chi^\dagger_{1/2}Y_{10}(\hat{\bs q})\right]\chi_{1/2}.
\en
Comparing this formula with Eq. (\ref{eq:tmelements}), we finally obtain
\eq
\tilde{{\cal A}}_{1/2}=-16\pi iE\sqrt{2}\frac{Q}{|{\bs k}|}S_{1+}.
\en
The two other cases in Eq. (\ref{eq:cases}) can be considered along the same lines. We get
\begin{eqnarray}
{\cal A}_{1/2}=-\frac{1}{2}(3E_{1+}+M_{1+})(-16\pi iE),\\\nonumber
{\cal A}_{3/2}=\frac{\sqrt{3}}{2}(E_{1+}-M_{1+})(-16\pi iE).
\end{eqnarray}

\end{document}